\documentclass[aps,reprint,prl,superscriptaddress,floats,floatfix]{revtex4-2}

\usepackage{amsmath}
\usepackage{amsfonts}
\usepackage{amssymb}
\usepackage{graphicx}
\usepackage{color}
\usepackage{tikz}
\usepackage{lipsum}
\usepackage{float}
\usepackage[colorlinks=true,linkcolor=blue,urlcolor=blue,citecolor=blue,pdfusetitle]{hyperref}
\usepackage{amstext}
\usepackage{latexsym}
\usepackage[running,mathlines]{lineno}
\usepackage{bbm}
\usepackage{verbatim}

\definecolor{dgreen}{rgb}{0.0, 0.5, 0.0}

\newcommand{\ket}[1]{|{#1}\rangle}
\newcommand{\bra}[1]{\langle{#1}|}

\newcommand{\ketbra}[2]{|{#1}\rangle\langle{#2}|}

\newcommand{\Tr}{\mathrm{Tr}}

\newcommand{\Zs}{Z_s}
\newcommand{\rhos}{\rho_s}
\newcommand{\varrhos}{\varrho_s}

\newcommand{\1}{\mathbbm{1}}
\newcommand{\tp}{\text{ . }}
\newcommand{\tc}{\text{ , }}

\newcommand{\tubingen}{Institut f\"{u}r Theoretische Physik, Eberhard Karls Universit\"{a}t T\"{u}bingen, Auf der Morgenstelle 14, 72076 T\"{u}bingen, Germany}

\newcommand{\Nottingham}{School of Physics and Astronomy and Centre for the Mathematics and Theoretical Physics of Quantum Non-Equilibrium Systems, The University of Nottingham, Nottingham, NG7 2RD, United Kingdom}

\begin{document}
	
\title{Large deviation full counting statistics in adiabatic open quantum dynamics}

\author{Paulo J. Paulino}
\email{paulo.paulino.souza96@gmail.com}
\affiliation{\tubingen}

\author{Igor Lesanovsky}
\affiliation{\tubingen}
\affiliation{\Nottingham}

\author{Federico Carollo}
\affiliation{\tubingen}

\date{\today}

\begin{abstract}
The state of an open quantum system undergoing an adiabatic process evolves by following the instantaneous stationary state of its time-dependent generator. This observation allows one to characterize, for a generic adiabatic evolution, the average dynamics of the open system. However,  information about fluctuations of dynamical observables, such as the number of photons emitted or the time-integrated stochastic entropy production in single experimental runs, requires controlling the whole spectrum of the generator and not only the stationary state. Here, we show how such information can be obtained in adiabatic open quantum dynamics by exploiting tools from large deviation theory. We prove an adiabatic theorem for deformed  generators, which allows us to encode, in a biased quantum state, the full counting statistics of generic time-integrated dynamical observables. 
We further compute the probability associated with  an arbitrary ``rare" time-history of the observable and derive a dynamics which realizes it in its typical behavior. Our results provide a way to  characterize and engineer adiabatic open quantum dynamics and to control their fluctuations. 
\end{abstract}

\maketitle

\begin{figure}[t]
    \centering
     \includegraphics[width=\linewidth,height=\textheight,keepaspectratio]{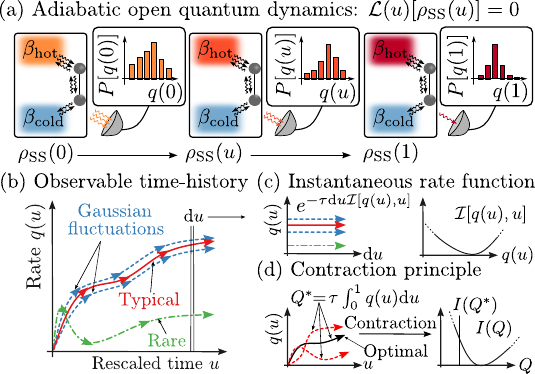}
     \caption{\textbf{Adiabatic open quantum dynamics and large deviations.} (a) During an adiabatic dynamics (total evolution time $\tau$), the system is described by the instantaneous steady-state $\rho_{\rm SS}(u)$ of the dynamical generator $\mathcal{L}(u)$, in the slow rescaled timescale $u=t/\tau$. (b) Sketch of different time-histories of the instantaneous ``rate" $q(u)$ of an observable of interest, e.g., the instantaneous photon-emission rate. We highlight the {\it typical} time-history,  two with small Gaussian fluctuations and a rare one displaying a {\it large deviation} from the typical value. (c) Due to the slow dynamics, the system spends a large amount of time in each rescaled time-interval ${\rm d}u$. This implies that the probability function for each instantaneous rate $q(u)$ [see illustration in panel (a)] obeys a large deviation principle. Combining them provides the probability functional for time-histories of the observable. (d) The full counting statistics of the time-integrated observable, $Q(\tau)=\tau \int_0^1{\rm d }u \, q(u)$, can be obtained from the optimal, i.e., most likely, trajectories providing the values of $Q(\tau)=Q^*$. }
    \label{fig:fig_1}
\end{figure}

\textit{Introduction.---}
Systems evolving adiabatically, i.e., via   slow driving protocols, find many applications in  physics. In closed quantum systems, adiabatic dynamics are characterized by the decoupled evolution of the Hamiltonian eigenvectors  \cite{born1928beweis,kato1950adiabatic, ballentine2014quantum, sarandy2004consistency, PhysRevLett.104.120401,berry1984, RevModPhys.82.1959, shapere1989geometric,teufel2003,teufel2020}, which is crucial for adiabatic quantum computation~\cite{RevModPhys.90.015002, PhysRevA.65.042308,PhysRevA.57.2403, PhysRevLett.99.070502, PhysRevLett.98.150402} and important experimental protocols such as stimulated Raman adiabatic passage~\cite{RevModPhys.89.015006}. In open quantum systems \cite{breuer2002theory, albash2012quantum,sarandy2005adiabatic, yi2007adiabatic,PhysRevLett.95.250503}, decoherence and dissipation typically impose a fundamental timescale in which this decoupled evolution can be observed, as explored in the context of optimal control~\cite{Alipour2020shortcutsto, yin2022shortcuts, PhysRevA.104.062421,PhysRevA.103.012206} and of noisy quantum computation~\cite{PhysRevLett.100.060503, PhysRevA.99.062320}. Adiabatic dynamics in open quantum systems occurs when the system state follows the instantaneous stationary state of its dynamical generator~\cite{DaviesSpohn1978,avron2012b,avron2012, PhysRevA.93.032118, PhysRevLett.119.050601,PhysRevA.95.042302} [see Fig.~\ref{fig:fig_1}{\color{blue}(a)}], as in the case of quasi-static thermodynamic  processes~\cite{callen1991thermodynamics,Deffner_quantum}.

Single realizations of open quantum dynamics, or quantum trajectories, are stochastic~\cite{wiseman2009quantum,RevModPhys.70.101}, which can manifest in the occurrence of quantum jumps, for instance related to  photon emissions~\cite{PhysRevA.98.042118,guerlin2007progressive, PhysRevA.97.022116,PhysRevResearch.2.033449, Manzano2022}.
For Markovian dynamics, the full counting statistics of jump-related observables, such as entropy production currents \cite{PhysRevLett.124.040602, PhysRevX.8.031037,Manzano2022}, can be obtained using {\it deformed dynamical generators}, introduced within the framework of large deviation theory \cite{TOUCHETTE20091, chetrite2015nonequilibrium, PhysRevLett.122.130605,PhysRevLett.109.250602,PhysRevLett.95.010601, lecomte2007thermodynamic, PhysRevLett.125.160601, BODINEAU2007540, PhysRevE.72.066110,PhysRevLett.104.160601,garrahan2016classical, PhysRevA.98.010103,cilluffo2021,cech2023}. 
However, much less is known about the characterization of dynamical fluctuations in open quantum dynamics with time-dependent generators 
\cite{PhysRevB.102.195409, liu2021auxiliary, PhysRevE.101.062144}, including the case of adiabatic processes.

In this work, we show how to fully characterize the counting statistics of jump-related observables in adiabatic open quantum dynamics [cf.~Fig.~\ref{fig:fig_1}{\color{blue}(a)}].
We prove an adiabatic theorem for deformed dynamical generators \cite{PhysRevLett.104.160601,PhysRevA.98.010103}, which allows us to demonstrate that, in these processes, the statistics of time-integrated observables assumes a large deviation form \cite{TOUCHETTE20091}. 
Furthermore, we show that adiabatic open quantum dynamics obey a so-called {\it temporal additivity principle} \cite{harris2009,harris2015,PhysRevE.106.054118,jack2020b}. That is,  the observables follow an instantaneous large deviation principle at all times [cf.~Fig.~\ref{fig:fig_1}{\color{blue}(b-c)}]. This fact opens up the possibility of deriving the probability of any time-history of the observable, 
see sketch in Fig.~\ref{fig:fig_1}{\color{blue}(b-c)}. Such a probability provides a higher level of description of dynamical fluctuations in the adiabatic process than what can be obtained from the full counting statistics of time-integrated observables. The latter can indeed be obtained from the former through a contraction principle \cite{TOUCHETTE20091},  [cf.~Fig.~\ref{fig:fig_1}{\color{blue}(c)}]. Finally, we construct an  auxiliary dynamics \cite{jack2010,PhysRevLett.104.160601,PhysRevA.98.010103} which can realize, as typical realization, any rare realization of the observable time-history in the original adiabatic process. Our findings (see Refs.~\cite{harris2009,harris2015,cavallaro2015,cavallaro2016,bertini2018,barato2018,PhysRevB.99.035437,sinitsyn2007berry, PhysRevLett.104.170601, hoppenau2013work,gu2023work, PhysRevLett.99.220408,PhysRevE.106.054118} for related results in classical dynamics) shed new insights on open quantum adiabatic processes and provide a powerful approach to control, even as a function of time, their fluctuating properties.
Our methods can be used for studying  fluctuations in adiabatic quantum machines~\cite{josefsson2018quantum, PhysRevResearch.2.033449, PhysRevE.99.042135}, both in or out of equilibrium, or for dissipative quantum computation~\cite{verstraete2009quantum, lin2013dissipative, PhysRevB.102.125112,PRXQuantum.4.010324}. 
\\

\textit{Open quantum dynamics.---}
We consider quantum systems whose dynamics is described by the master equation  $\dot{\rho}(t)= \mathcal{L}(t)[\rho(t)]$, with time-dependent generator 
\begin{equation}\label{eq:linbladEq}
    \mathcal{L}(t)[\rho]=-i\left[\Tilde{H}(t) \rho - \rho \Tilde{H}^\dag(t) \right] + \sum_j \mathcal{J}_j(t)[ \rho] \, .
\end{equation}
Here, $\Tilde{H}(t)= H(t) -(i/2)\sum_j  J_j^\dag(t) J_j(t) $ is the effective Hamiltonian \cite{RevModPhys.70.101},  $\mathcal{J}_j(t)[\rho] = J_j(t) \rho J_j^{\dag}(t)$, with $J_j(t)$ being the jump operators.
The above equation generates the evolution of the system state, $\rho(t)$, averaged over all possible realizations of the system-environment interaction \cite{breuer2002theory,wiseman2009quantum}. Single dynamical realizations are instead described by quantum jump trajectories  \cite{PhysRevA.35.198, PhysRevLett.104.160601, PhysRevA.98.010103}, generated by the stochastic process 
\begin{equation}
    {\rm d}  \psi(t)\!=\! \mathcal{B}(t)[\psi(t)]   {\rm d}   t +\! \sum_j\!\!\left(\!\frac{\mathcal{J}_j(t)[\psi(t)]}{\Tr\left(\mathcal{J}_j(t)[\psi(t)]\right)} - \psi(t)\!\right) \! {\rm d}n_j(t),
\end{equation}
which evolves pure quantum states $\psi=\ket{\psi}\bra{\psi}$. Here, ${\rm d}  \psi(t)$ is the state increment while ${\rm d}n_j(t)$ are Poisson increments, which can only take the value $0$ or $1$ with average value $\mathbb{E}_{\psi(t)=\psi}[{\rm d}n_j(t)]={\rm d}t \, \Tr\left(\mathcal{J}_j(t)[\psi]\right)$ \cite{breuer2002theory, wiseman2009quantum}, where $\mathbb{E}_{\psi(t)=\psi}$ denotes the expectation over the process conditional to the system being in $\psi$ at time $t$. When a Poisson increment is equal to $1$, the state undergoes a jump associated with the corresponding $\mathcal{J}_j(t)$. When all increments are zero, the system evolves continuously through the map 
\begin{equation*}
    \mathcal{B}(t)[\psi] = -i \Tilde{H}(t) \psi +i \psi \Tilde{H}^\dag(t) - \psi \Tr[-i \Tilde{H}(t)\psi +i \psi \Tilde{H}^\dag(t) ] \tp
\end{equation*}

A generic time-integrated observable associated with quantum-jump events can thus be defined as 
\begin{equation}\label{eq:generalized_activity}
    Q(t) =  \sum_j \int_0^t f_j(v){\rm d}n_j(v) \tp
\end{equation}
When $f_j(v)=1$ $\forall j$, $Q(t)$ equals the total number of jumps occurred during a trajectory. For other choices, it is instead related to, for instance, stochastic heat or entropy production in thermal machines~\cite{PhysRevE.85.031110, PhysRevX.8.031037, PhysRevResearch.2.033449, Manzano2022}. 
To characterize the properties of this  observable, 
it is convenient to work with its moment generating function, defined as $ \Zs(t) = \mathbb{E}[e^{-s  Q(t)}]$ through the field $s$, which is conjugate to the observable.  As shown in the Supplemental Material (SM)~\cite{Supp}, the moment generating function can be computed as $ \Zs(t) = {\rm Tr}(\rho_s(t))$, where $\dot{\rho}_{s}(t) = \mathcal{L}_{s}(t)[\rhos(t)]$ and with $\mathcal{L}_s(t)$ being the  deformed dynamical generator \cite{PhysRevLett.104.160601, PhysRevLett.111.120601}
\begin{equation}\label{eq:tilted_Lindbladian}
    \mathcal{L}_s(t)[\rho] = \mathcal{L}(t)[\rho] + \sum\limits_j(e^{-s f_j(t)}-1) \mathcal{J}_j(t)[\rho] \tp
\end{equation}
For time-independent deformed generators and large evolution times $\tau$, $Z_s(\tau)$ obeys a large deviation principle,  $Z_s(\tau)\approx e^{\tau\theta_s}$ with $\theta_s$ being the scaled cumulant generating function of $Q(\tau)$. In such a time-independent framework, $\theta_s$ coincides with the dominant eigenvalue of $\mathcal{L}_s$ \cite{TOUCHETTE20091, PhysRevLett.104.160601} and  fully characterizes the probability $P[Q(\tau)=Q^*]$. This also takes a large deviation form $P[Q(\tau)=Q^*]\approx e^{-\tau I(Q^*/\tau)}$, with rate function given by the Legendre-Fenchel transform $I(x) = \sup_{s \in \mathbb{R}}\{-s x- \theta_{s}\}$~\cite{TOUCHETTE20091}.
In what follows, we derive the behavior of $Z_s(\tau)$ for the case of adiabatic open quantum dynamics.
To this end, we consider that $\dot{Z}_s(t)={\rm Tr}\left(\mathcal{L}_s(t)[\varrho_s(t)]\right)Z_s(t)$, where $\varrhos(t) = \rhos(t)/\Tr\left(\rhos(t)\right)$, which we can use to express the moment generating function as 
\begin{equation}\label{eq:Partition_function}
    \Zs(\tau) = e^{\int_0^\tau \Tr\left(\mathcal{L}_s(t) \left[\varrhos(t)\right]\right) {\rm d}t} \, .
\end{equation}
As we show below, this expression allows us to write $\theta_s$ in terms of the instantaneous dominant eigenvalues of $\mathcal{L}_s(t)$.\\

\textit{Adiabatic theorem for deformed generators.---}
We consider $\mathcal{L}_s(t)$ to vary on the slow timescale $u=t/\tau$, with $\tau$ being the total evolution time and we assume it to be diagonalizable with right and left eigenmatrices, $r^m_s(t)$ and $\ell^m_s(t)$. These are such that $\mathcal{L}_s(t)[r_s^m(t)]=\lambda_s^m(t) r^m_s(t)$ and $\mathcal{L}_s^*(t)[\ell_s^m(t)]=\lambda_s^m(t)\ell_s^m(t)$, where $\lambda_s^m(t)$ are the instantaneous eigenvalues of $\mathcal{L}_s(t)$  and $\mathcal{L}_s^*(t)$ is the dual generator acting on operators. We consider the dominant eigenvalue $\lambda_s^0(t)$ to be  unique (and thus real), so that $\lambda_s^0(t) > {\rm Re}\{\lambda_s^m(t)\}$, for $m\ge1$. With these definitions, our adiabatic condition reads  (C1) $\|\dot{r}_s^m(t)\|,\|\dot{\ell}_s^m(t)\|,|\dot{\lambda}_s^m(t)|\sim 1/\tau$, encoding that the generator varies slowly for large $\tau$. Our second assumption is related to the uniqueness of the dominant eigenvalue $\lambda_s^0(t)$ and is conveniently expressed as the existence of a finite gap $\Delta$ for all times: (C2) $\Delta:=\inf_{m>0,\forall t}\left\{|\lambda_s^0(t)-{\rm Re}\{\lambda_s^m(t)\}|\right\}>0$. 

Given the two assumptions above, we prove in the  SM~\cite{Supp} that, within the rescaled slow timescale $u=t/\tau$, 
\begin{equation}
\lim_{\tau\to\infty}\varrho_s(u)=r_s^0(u)\, , \quad  \, 0<u\le 1.
\label{ad-theorem}
\end{equation}
Note that, with a slight abuse of notation we denote the dependence on the slow timescale $u$ in the same way as that on the original timescale $t$. Eq.~\eqref{ad-theorem} shows that under the evolution with the deformed dynamical generator the normalized state $\varrho_s(u)$ follows the path of the instantaneous dominant right eigenmatrix of $\mathcal{L}_s(t)$. This result thus extends the adiabatic theorem for open quantum systems~\cite{DaviesSpohn1978,avron2012b,avron2012, PhysRevA.93.032118, PhysRevLett.119.050601,PhysRevA.95.042302} to deformed dynamical generators and includes, for $s=0$, the case of completely generic open quantum dynamics satisfying conditions (C1-C2). Importantly, controlling the evolution of the state under the deformed dynamical generator, as in our result, does only provide information about the stationary state ($s=0$ case) as in usual adiabatic theorems, but also encodes information (for $s\neq0$) about the whole spectrum of excitations of the  generator of the adiabatic open quantum dynamics. Finally, through our approach Eq.~\eqref{ad-theorem} can be shown to be valid independently on the initial state of the system.

As a consequence of Eq.~\eqref{ad-theorem}, the moment generating function $Z_s(\tau)$ in Eq.~\eqref{eq:Partition_function} obeys a large deviation principle, in the limit $\tau \to\infty$, with scaled cumulant generating function given by  
\begin{equation}
    \theta_s^{\rm ad}=\int_0^1 \lambda_s^0(u) {\rm d }u\, .
    \label{theta-adiabatic}
\end{equation}
As such, the statistics of $Q(\tau)$ also obeys a large deviation principle \cite{TOUCHETTE20091}, characterized by the function $I(Q/\tau)$, obtained as the Legendre-Fenchel transform of $\theta_s^{\rm ad}$. Interestingly,  Eq.~\eqref{theta-adiabatic}  remains valid also in the case of  degenerate dominant eigenvalues $\lambda_s^0(u)$ \cite{Supp}.

\begin{figure}[t]
    \centering
     \includegraphics[width=\linewidth,height=\textheight,keepaspectratio]{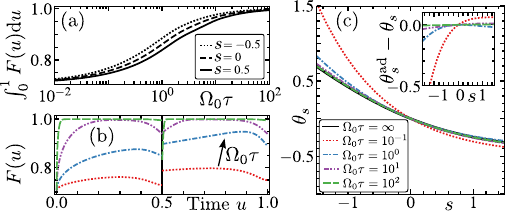}
    \caption{\textbf{Driven two-level atom.}  (a) Time-integrated fidelity $\int_0^1 F(u) {\rm d}u$, with  $F(u)= \Tr(\sqrt{{\varrho_s^{1/2}(u)}r_s^0(u){\varrho^{1/2}_s(u)}}$, for $s=-0.5$ (dotted line), $s=0$ (dashed line), and $s=0.5$ (solid line). (b) The instantaneous fidelity $F(u)$ for different values of $\Omega_0\tau$, see legend in panel (c). The parameters are  $\Omega_0=\gamma$ and $s=-0.5$. 
    (c) Scaled cumulant generating function for the activity. The black solid line corresponds to $\theta_s^\text{ad}$, while the other lines give the function $\theta_s$ for different values of $\Omega_0\tau$. The inset shows how $\theta_s$ approaches $\theta_s^\text{ad}$ as  $\Omega_0\tau$ is increased.}
    \label{fig:Fig2}
\end{figure}

To benchmark these results, we consider a resonantly driven two-level atom, with excited state $\ket{e}$, ground state $\ket{g}$ and Hamiltonian $H(t)= \Omega(t/\tau) (\sigma_++\sigma_-)$, where we defined $\sigma_-=\sigma_+^\dagger=\ket{g}\bra{e}$. We assume  $\Omega(u)=\Omega_0\cos(u\pi)$ for $u<1/2$ and $\Omega_0\sin(u\pi)$ for $u\ge1/2$.
The atom emits photons, which is described by the jump operator  $J=\sqrt{\gamma} \sigma_-$, where $\gamma$ is the emission rate. 
We focus on the activity, i.e., the total number of quantum jumps, $A(\tau)=\int_0^\tau {\rm d}n(t)$ \cite{PhysRevLett.104.160601}.
Fig.~\ref{fig:Fig2}{\color{blue}(a)} shows the time-averaged fidelity between $r_s^0(u)$ and $\varrhos(u)$ as a function of the total time  $\tau$. The inset displays the fidelity as a function of the rescaled time $u$, for increasing $\tau$. The results confirm our theorem as well as the convergence of the scaled cumulant generating function to the one in Eq.~\eqref{theta-adiabatic}, as $\tau\to\infty$ [see Fig.~\ref{fig:Fig2}{\color{blue}(b)}]. They further show that our findings remain valid in the case of piecewise-differentiable dynamical parameters. \\

\textit{Time-history of the observable.---}
The scaled cumulant generating function in Eq.~\eqref{theta-adiabatic}, together with its Legendre-Fenchel transform, characterizes the time-integrated observable $Q(\tau)$ during an adiabatic process.
However, it is also relevant  to characterize the probability of the different time-histories of the observable  [cf.~Fig.~\ref{fig:fig_1}{\color{blue}(b)}] realizing different values of $Q(\tau)$. 
To arrive at such a higher level of description of the process, we observe that, due to the adiabatic nature of the open quantum dynamics, the  system spends an infinite amount of time in each of the infinitesimal (rescaled) time-intervals ${\rm d}u$.  
For each ${\rm d}u$, it is possible to define a coarse-grained instantaneous rate $q(u)$, representing the time-averaged value of the observable at the rescaled time $u$ (see Ref.~\cite{Supp} for  details). 
We can thus write $Q(\tau)=\tau\int_0^1 q(u) {\rm d}u$, where $\{q(u)\}$ is a (stochastic) time-history of the observable rate, as illustrated in Fig.~\ref{fig:fig_1}{\color{blue}(b)}. 
Discretizing time and considering that each $q(u)$ obeys an independent large deviation principle, we have that the probability over time-histories is given by $P[\{q(u)\}]\approx \prod_u P[q(u)]$ and, in the continuous-time limit, 
\begin{equation}\label{eq:records_prob}
   P[\{q(u)\}] \asymp e^{-\tau \varphi[\{q(u)\}]} \tc  \varphi[\{q(u)\}]\!=\! \int_0^1 \mathcal{I} [q(u), u]{\rm d}u \tc
\end{equation}
where $\mathcal{I} [q(u), u]$ is the instantaneous large deviation function of $q(u)$, i.e., the Legendre-Fenchel transform of $\lambda_s^0(u)$ \cite{zia2009making}. From a physical perspective, we expect time-histories $\{q(u)\}$ to be sufficiently regular, e.g., piecewise analytic functions of time. 
Essentially, this shows that adiabatic open quantum dynamics obey the so-called temporal additivity principle introduced in Ref.~\cite{harris2009}. (See the SM for the formal proof~\cite{Supp}).

The functional in Eq.~\eqref{eq:records_prob} contains the full  information about time-histories of the observable rate $\{q(u)\}$ and, thus, a complete description of fluctuations at the rescaled timescale  $u$. The typical time-history is the one minimizing the functional $\varphi$, that is, the one passing through the minima of the instantaneous rate functions $\mathcal{I} [q(u), u]$. 
The functional $\varphi$ can further be used to derive the statistics of any observable constructed from the time-history $\{q(u)\}$. An example is again the time-integrated observable $Q(\tau)$, whose functional  $I$ can be retrieved, via a contraction principle \cite{TOUCHETTE20091}, as 
\begin{equation}
    \label{eq:contraction_principle}
I(x)=\inf_{\forall \{q(u)\}: \, x=\int_0^1q(u){\rm d}u } \varphi[\{q(u)\}].
\end{equation}
Physically, this means that the probability of observing $Q=Q^*$ is equal to the probability of the most likely time-history $\{q^*(u)\}$ providing value of the time-integrated observable [cf.~Fig.~\ref{fig:fig_1}{\color{blue}(d)}]. 

While the general derivation of the contraction in Eq.~\eqref{eq:contraction_principle} is provided in the SM \cite{Supp}, we discuss it here using the example of the two-level atom, setting for convenience $\gamma(u)=4\Omega(u)$  \cite{PhysRevLett.104.160601}. In this case, we find  $\lambda_s^0(u) = 2\Omega(u)\left(e^{-s/3}-1\right) $ and $ \mathcal{I}[a(u), u] = 3[a(u)\log (a(u)/a_0(u)) - (a(u)-a_0(u))]$, where $a_0(u)=(2/3)\Omega(u)$ is the typical time-history of the activity rate. To compute the minimization in Eq.~\eqref{eq:contraction_principle}, we perform a functional derivative and set it to zero. This results in  $a^*(u) = a_0(u)e^{-\mu/3}$ where $\mu$ is a Lagrange multiplier introduced to enforce the constraint in Eq.~\eqref{eq:contraction_principle}. Integrating $a^*(u)$ over time, we find $A^*=A_0e^{-\mu/3}$ which fixes the Lagrange multiplier to $\mu^* =3\log (A_0/A^*)$ with $A_0$ being the typical value of the time-integrated observable $A(\tau)$. Substituting this information  into the functional $\varphi$ [cf.~Eq.~\eqref{eq:records_prob}], we find $I(A^*/\tau)=3[(A^*/\tau)\log A^*/A_0  -(A^*-A_0)/\tau]$, which is the same result one gets by calculating the Legendre-transform of $\theta_s^{\rm ad}$ given in Eq.~\eqref{theta-adiabatic} \cite{PhysRevLett.104.160601}.

\begin{figure}[t]
    \centering
     \includegraphics[width=\linewidth,height=\textheight,keepaspectratio]{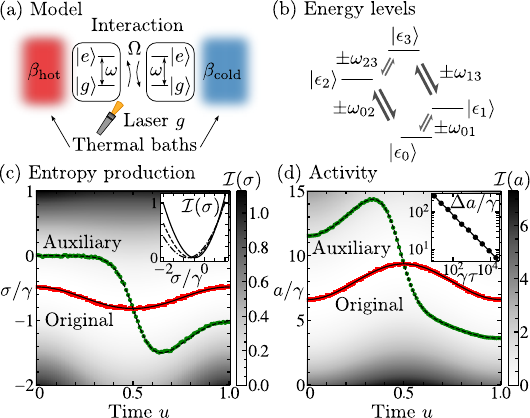}
    \caption{\textbf{Two-atom system.} (a) Two qubits (bare energy $\omega$) interact by exchanging excitations at rate $\Omega$. 
    The qubit on the left (right) is in  contact with a hot (cold) bath, with inverse temperature $\beta_{\rm hot}$ ($\beta_{\rm cold}$). The system is also subject to a laser driving with Rabi frequency $g$. 
    (b)
    Transitions implemented by the jump operators, with $\omega_{ij}$ representing the energy exchanges with the thermal baths given by  $\omega_{02}=\omega_{13}=(\omega+\Omega)$ and $\omega_{01}=\omega_{23}=(\omega-\Omega)$. (c-d) Instantaneous rate functions $\mathcal{I}(q(u),u)$ for the entropy production [$\sigma(u)$] and the activity rates [$a(u)$] as a function of time, respectively.  The black solid line shows the typical time-history of the rates while the dashed line a rare one. Red squares are estimates obtained by running a (typical) quantum trajectory evaluated at $s=0$, while the green bullets a quantum trajectory with the auxiliary system in Eq.~\eqref{Doob_H}, for $s(u) = (1/2)\tanh(10u -5)$. The total evolution is $ \gamma\tau = 2 \times 10^6$. The inset in (c) shows the instantaneous rate function for the entropy production for three different times: $u = 0$ (solid line), $u=1/4$ (dashed line), and $u=1/2$ (dot-dashed line). The inset in (d) shows the error between the activity rate estimated through a  quantum trajectory, $a(u)$, and the expected  behavior, $a_0(u)$, in the adiabatic limit, $\Delta a = \int_0^\tau |a(u) - a_0(u)|\mathrm{d} u$, as a function of $\gamma\tau$.  The parameters are $g=2\Omega=\omega=\gamma$ and the temperatures are $\beta_{\rm hot}^0=\beta_{\rm cold}/2=1$, in units of $1/\omega$.}
    \label{fig:Fig3}
\end{figure}

The functional $\varphi$ is formally derived as the Legendre-Fenchel transform of the scaled cumulant generating ``functional"  $\Theta[\{s(u)\}]$, associated with a time dependent field $s(u)$ \cite{Supp}. The function  $\Theta[\{s(u)\}]$ is obtained as in Eq.~\eqref{theta-adiabatic} from the eigenvalues of the deformed operator in Eq.~\eqref{eq:tilted_Lindbladian} defined with a time-dependent $s(u)$. The knowledge of such a deformed operator gives us a handle to manipulate on-demand time-histories of the observable: by choosing $s^*(u)$ such that $-\delta \Theta[\{s^*(u)\}]/\delta s^*(u)=q^*(u)$, we can indeed define a suitable open quantum dynamics which produces, as typical, the rare time-history $\{q^*(u)\}$ of the original process \cite{jack2010,PhysRevLett.104.160601,PhysRevA.98.010103}. Such an auxiliary dynamics is given by \cite{Supp} 
\begin{equation}
\begin{split}
    \label{Doob_H}
    &H^{\rm A}(u)=\frac{1}{2}\Bigl\{\left[\ell_{s^*(u)}^{0}(u)\right]^{1/2}\tilde{H}(u) \left[\ell_{s^*(u)}^{0}(u)\right]^{-1/2} + \text{H.c.}\Bigr\}\, ,\\
    &{J}^{\rm A}_j(u)=e^{-\frac{s^*(u)}{2}f(u)} \left[\ell_{s^*(u)}^{0}(u)\right]^{1/2} J_j(u) \left[\ell_{s^*(u)}^{0}(u)\right]^{-1/2}\, , 
    \end{split}
\end{equation}
where $\ell_{s^*(u)}^{0}(u)$ is the dominant left eigenmatrix of the deformed operator with time-dependent  field $s^*(u)$. \\ 

\textit{Stochastic entropy production.---} As an application of our results, we consider a system composed by two two-level atoms attached to different thermal baths [cf.~Fig.~\ref{fig:Fig3}{\color{blue}(a)}]. The system Hamiltonian is $H=\omega (\sigma_e^{\rm hot}+\sigma_e^{\rm cold})+\Omega(\sigma_-^{\rm hot}\sigma_+^{\rm cold}+\text{H.c.})$, with $\sigma_{e}=\ket{e}\bra{e}$ and where the superscripts hot and cold indicate the atom in contact with the corresponding thermal reservoir  [cf.~Fig.~\ref{fig:Fig3}{\color{blue}(a)}]. 
The dynamics is governed by a time-dependent Lindblad generator derived via a weak coupling of the system with the thermal baths~\cite{breuer2002theory,Supp}. 
 The jump operators thus read  $J_{ij}^b = \sqrt{\gamma}\sqrt{(\omega_{ji}/\omega)^3 N_b(\omega_{ji})}\ketbra{\epsilon_i}{\epsilon_j}$, where $b$ indexes the baths, $\gamma$ is a rate,  $N_b(\omega)=1/(e^{\beta_b\omega}-1)$, and $\omega_{ij}=\epsilon_j -\epsilon_i$ is the difference between the energies of the eigenstates $\ket{\epsilon_j}$ and $\ket{\epsilon_i}$ of $H$~\cite{breuer2002theory}. 
They implement transitions between the eigenstates $|\epsilon_i\rangle$ and $|\epsilon_j\rangle$ of $H$ [cf.~Fig.~\ref{fig:Fig3}{\color{blue}(a-b)}]. 
In addition, we include a phenomenological laser driving term, $H_{\text{laser}}=g(\sigma_x^{\text{hot}} + \sigma_x^{\text{cold}})$ \cite{PhysRevLett.124.170602,nill2022,paulino2022nonequilibrium}.
The stochastic entropy production associated with any quantum jump in the dynamics is defined as the energy exchanged with the corresponding thermal bath responsible for the transition, $\pm \Delta \sigma^b_{ij} = \pm \omega_{ij}\beta_b$~\cite{PhysRevX.8.031037, PhysRevResearch.2.033449,  Manzano2022}. The time-integrated entropy flow from the hot bath to the cold one can thus be defined as $\Sigma (\tau)= \sum_{i,j, b} \int_0^\tau \beta_b(t) \omega_{ij} {\rm d} n_{ij}^b(t) $,
where ${\rm d} n_{ij}^b$ are the increments associated with the different jump operators~\cite{PhysRevX.8.031037}. We also consider the activity $A(\tau) = \sum_{i,j,b} \int_0^\tau {\rm d} n_{ij}^b(t) $. The inverse temperature of the hot bath follows the protocol  $\beta_\text{hot} (t)= \beta_\text{hot}^0/(1+(1/2)\sin^2(t\pi/\tau))$.

Figs.~\ref{fig:Fig3}{\color{blue}(c-d)} show the instantaneous rate functions $\mathcal{I}$, for the two observables. 
The rate function for the entropy current displays  a symmetry \cite{marcantoni2020,marcantoni2021} related to the existence of entropy fluctuation relations at all times $t$ [see inset of Fig.~\ref{fig:Fig3}{\color{blue}(c)}]. In Figs.~\ref{fig:Fig3}{\color{blue}(c-d)}, we further provide numerical results 
from quantum trajectories for both the original dynamics and an auxiliary one displaying a rare realization of the observables.  \\

\textit{Discussion.---} 
We have derived a complete statistical characterization for open quantum systems in the adiabatic regime.
Our analysis can be extended to different quantum stochastic processes, such as diffusive quantum trajectories associated with homodyne-detection experiments~\cite{gardiner2004quantum}. 
It would be interesting to explore whether the auxiliary quantum dynamics derived here can be exploited to control the performance of (adiabatic) quantum machines~\cite{PhysRevA.94.042132, Kieferova_2014, coulamy2016energetic}.
With regards to applications in adiabatic quantum computing~\cite{PhysRevLett.95.250503, PhysRevA.102.052215}, it would be important to generalize our analysis to characterize the full counting statistics of  state-dependent observables \cite{PhysRevLett.122.130605}, such as the fidelity. This would require the application of numerical schemes as in Ref.~\cite{PhysRevE.102.030104} or the derivation of a level 2.5 formalism for adiabatic open quantum dynamics \cite{PhysRevLett.122.130605,carolloJStatPhys}. \\

\textit{Acknowledgements.--} The codes used to produce the numerical results of this paper are available on Github \cite{Github}. We are grateful to Stefan Teufel and Tom Wessel for useful discussions on adiabatic theorems and to Robert L. Jack for drawing our attention to related literature.  We acknowledge funding from the Deutsche Forschungsgemeinschaft (DFG, German Research Foundation) under Project No. 435696605 and through the Research Units FOR 5413/1, Grant No. 465199066 and FOR 5522/1, Grant No. 499180199. This project has also received funding from the European Union’s Horizon Europe research and innovation program under Grant Agreement No. 101046968 (BRISQ). F.C.~is indebted to the Baden-W\"urttemberg Stiftung for the financial support of this research project by the Eliteprogramme for Postdocs.\\

\bibliography{refs}

\newpage

\renewcommand\thesection{S\arabic{section}}
\renewcommand\theequation{S\arabic{equation}}
\renewcommand\thefigure{S\arabic{figure}}
\setcounter{equation}{0}
\setcounter{figure}{0}

\onecolumngrid

\newpage

\setcounter{page}{1}

\begin{center}
{\Large SUPPLEMENTAL MATERIAL}
\end{center}
\begin{center}
\vspace{0.8cm}
{\Large Large deviation full counting statistics in adiabatic open quantum dynamics}
\end{center}

\begin{center}
Paulo J. Paulino$^1$, Igor Lesanovsky$^{1,2}$ and Federico Carollo$^{1}$ 
\end{center}
\begin{center}
$^1${\em Institut f\"{u}r Theoretische Physik,  Universit\"{a}t T\"{u}bingen, Auf der Morgenstelle 14, 72076 T\"{u}bingen, Germany,}\\
{\em Auf der Morgenstelle 14, 72076 T\"ubingen, Germany}\\
$^2${\em School of Physics and Astronomy and Centre for the Mathematics}\\
{\em and Theoretical Physics of Quantum Non-Equilibrium Systems,}\\
{\em  The University of Nottingham, Nottingham, NG7 2RD, United Kingdom}\\

\end{center}

\section{I. Derivation of the deformed dynamical generator}

In this Section,  we derive the deformed dynamical generator $\mathcal{L}_{s}(t)$, which is given in Eq.~\eqref{eq:tilted_Lindbladian} of the main text. 
We start by defining the time-integrated observable 
\begin{equation}
    Q(t) =  \sum_j \int_0^t  f_j(v) \mathrm{d}n_j(v)\, ,
\end{equation}
where $\mathrm{d}n_j(v)$ is a Poisson increment associated with a jump through the $j$-th channel, at time $t$. The noises are such that $\mathrm{d}n_j^2(t)= \mathrm{d}n_j(t)$.  

To calculate the moment generating function,   we have to evaluate $Z_{s}(t) = \mathbb{E}[K(t)]$, where $K(t) = e^{-s Q(t)}$. 
To this end, it is convenient to turn the problem into finding the evolution of the biased density matrix 
\begin{equation}
    \rho_{s}(t) = \mathbb{E}[K(t) \psi(t)] = \int\limits_0^t P[\psi(t)=\psi] \mathbb{E}\big\rvert_{\psi(t)=\psi}[K(t)\psi(t)] \mathrm{d}\psi =  \int\limits_0^t P_{s}[\psi(t)=\psi]\psi \mathrm{d}\psi \tc
\end{equation}
where $P_{s}[\psi(t)=\psi]= P[\psi(t)=\psi] \mathbb{E}\big\rvert_{\psi(t)=\psi}[K(t)]$. The moment generating function can be retrieved using the fact that $Z_s(t)={\rm Tr}(\rho_s(t))$. 

In order to find the map that generates the dynamics of the deformed density matrix, $\rho_{s}(t)$, we consider the differential equation 
\begin{equation}
    \mathrm{d}\rho_{s}(t) = \int\limits_0^t P[\psi(t)=\psi] \mathbb{E}\big\rvert_{\psi(t)=\psi}[\mathrm{d}(K(t)\psi(t))] \mathrm{d}\psi\tp
\end{equation} 
The differential inside the integral can be written as the sum of three different components, since 
\begin{equation}
\label{diff-terms}
    \mathrm{d}(K(t)\psi(t))=\mathrm{d}(K(t))\psi(t)+K(t) \mathrm{d}(\psi(t))+ \mathrm{d}(K(t))\mathrm{d}(\psi(t))\, ,
\end{equation}
and we evaluate each of the above increments independently. 
For the term $K(t)$, we have 
\begin{equation}
    K(t+\mathrm{d}t) =  e^{-\sum_j\int_0^{t+\mathrm{d}t}s(v)f_j(v)\mathrm{d}n_j(v)} = K(t) e^{-s(t)\sum_j f_j(t)\mathrm{d}n_j(t)} \tp
\end{equation}
We now can explicitly calculate the exponential exploiting the relation $\mathrm{d}n_i(t)\mathrm{d}n_j(t)=\mathrm{d}n_i(t)\delta_{ij}$, where $\delta_{ij}$ is a Kronecker delta. With this we can write  
\begin{equation}
   \mathrm{d}K(t) = K(t+\mathrm{d} t) - K(t) = K(t)\sum_j\mathrm{d}n_j(t)(e^{-s(t)f_j(t)} - 1)  \, ,
\end{equation}
which allows us to calculate the expectation of the first term appearing on the right hand side of Eq.~\eqref{diff-terms} as 
\begin{equation}
 \mathbb{E}\big\rvert_{\psi(t)=\psi}[\mathrm{d}K(t)\psi(t)] = \mathbb{E}\big\rvert_{\psi(t)=\psi}[K(t)]\psi\sum\limits_j\left(e^{-sf_j(t)}-1\right)\Tr\Bigl\{\mathcal{J}_j[\psi]\Bigr\} \mathrm{d}t \tp 
\end{equation}

The stochastic equation for the pure state $\psi(t)$ is given by 
\begin{equation}
    {\rm d}  \psi(t)\!=\! \mathcal{B}(t)[\psi(t)]   {\rm d}   t + \sum_j\!\!\left(\frac{\mathcal{J}_j(t)[\psi(t)]}{\Tr\left(\mathcal{J}_j(t)[\psi(t)]\right)} - \psi(t)\right) \! {\rm d}n_j(t) \, ,
\end{equation}
where  $\Tilde{H}(t)= H(t) -(i/2)\sum_j J_j^\dag(t) J_j(t) $ is the effective Hamiltonian.
As defined in the main text, we have $\mathcal{B}(t)[\psi] = -i \Tilde{H}(t) \psi +i \psi \Tilde{H}^\dag(t) - \psi \Tr[-i \Tilde{H}(t)\psi +i \psi \Tilde{H}^\dag(t) ] $, and  $\mathcal{J}_j(t)[\rho] =  J_j(t) \rho J_j^{\dag}(t)$, with $J_j(t)$ being the jump operators.
Therefore, we have
\begin{equation}
 \mathbb{E}\big\rvert_{\psi(t)=\psi}[K(t)\mathrm{d}\psi(t)]=\mathbb{E}\big\rvert_{\psi(t)=\psi}[K(t)]\mathcal{L}[\psi]\mathrm{d}t \tp
\end{equation}

For the last term, $\mathrm{d}K(t) \mathrm{d}\psi(t)$, we further consider that $\mathrm{d}t \mathrm{d}n_j(t)$ is of higher-order in  $\mathrm{d}t$ and we obtain
\begin{equation}
    \mathrm{d}K(t)\mathrm{d}\psi(t) = K(t) \sum\limits_j\left(\frac{\mathcal{J}_j(t)[\psi(t)]}{\Tr\left(\mathcal{J}_j(t)[\psi(t)]\right)} - \psi(t)\right) \left(e^{-sf_j(t)} - 1\right){\rm d}n_j(t) \tp
\end{equation}
By taking the expectation value, we find 
\begin{equation}
    \mathbb{E}\big\rvert_{\psi(t)=\psi}[\mathrm{d}K(t)\mathrm{d}\psi(t)] = \mathbb{E}\big\rvert_{\psi(t)=\psi}[K(t)]\sum\limits_j\left(\mathcal{J}_j[\psi] - \psi \Tr\Bigl\{\mathcal{J}_j[\psi]\Bigr\}\right)\left(e^{-sf_j(t)}-1\right)\mathrm{d}t  \tp
\end{equation}
Summing all three contributions we find
\begin{equation}
    \mathbb{E}\big\rvert_{\psi(t)=\psi}[\mathrm{d}(K(t)\psi(t))] = \mathbb{E}\big\rvert_{\psi(t)=\psi}[K(t)]\left(\mathcal{B}[\psi] + \sum\limits_je^{-sf_j(t)}\mathcal{J}_j[\psi]\right)\mathrm{d}t \tp 
\end{equation}
To conclude the derivation, we note that 
\begin{equation}
    \mathrm{d}\rho_s(t) =  \int\limits_0^t P[\psi(t)=\psi] \mathbb{E}\big\rvert_{\psi(t)=\psi}[\mathrm{d}(K(t)\psi(t))] \mathrm{d}\psi =  \int\limits_0^t P[\psi(t)=\psi] \mathbb{E}\big\rvert_{\psi(t)=\psi}[K(t)] \mathcal{L}_{s}(t)[\psi] \mathrm{d}\psi = \mathcal{L}_{s}(t)[\rho_s(t)] \tc
\end{equation}
where 
\begin{equation}
       \mathcal{L}_{s}(t)[\rho] = -i\left[\tilde{H}(t) \rho - \rho \tilde{H}^\dag(t) \right] + \sum_j e^{-s f_j(t)} J_j(t) \rho  J_j^\dag(t) \tp 
\end{equation}

\section{II. Adiabatic Theorem for deformed dynamical generators}

In this Section, we provide a proof for the result presented in Eq.~\eqref{ad-theorem} in the main text. Assuming that the deformed generator $\mathcal{L}_s(t)$ can be diagonalized, we can write the deformed state of the system at time $t$, $\rho_s(t)$, as  
\begin{equation}
    \label{decomp}
    \rho_s(t) = \sum_m c_s^m(t) r_s^m(t)\, .
\end{equation}
Here, $c_s^m(t)={\rm Tr}(\ell_s^m(t) \rho_s(t))$ and $\ell_s^m(t),r_s^m(t)$ are the left and right eigenmatrices of the instantaneous generator $\mathcal{L}_s(t)$. The first step of the proof is to determine the time evolution of the coefficients $c_s^m(t)$. To this end, we take the time derivative of $\rho_s(t)$ which gives 
\begin{equation}
    \label{t-der}
    \dot{\rho}_s(t) = \sum_m \left[\dot{c}_s^m(t) r_s^m(t) +{c}_s^m(t) \dot{r}_s^m(t) \right]=\sum_ m \lambda_s^m(t) c_s^m(t) r_s^m(t)\, ,
\end{equation}
where the second equality comes from acting on $\rho_s(t)$ with the deformed generator $\mathcal{L}_s(t)$. To find an equation for the coefficients $c_s^m(t)$, we ``project" the above equation onto $\ell_s^k(t)$, exploiting the orthogonality relation ${\Tr }(\ell_s^k(t) r_s^m(t))=\delta_{km}$, to get 
\begin{equation}
    \label{coeff-der}
    \dot{c}_s^k(t) = \lambda_s^k(t) c_s^k(t) - \sum_m c_s^m(t) {\rm Tr}\left(\ell_s^k(t) \dot{r}_s^m(t)\right)\, .
\end{equation}
Introducing the vector ${\bf c}_s(t)=[c_s^0(t),c_s^1(t), \dots ]^T$, we can recast the above system of differential equations as 
\begin{equation}
    \label{coeff-der-vec}
\dot{\bf c}_s(t)=\left[\Lambda(t)+\frac{1}{\tau} M(t)\right]{\bf c}_s(t)\, ,
\end{equation}
where $\Lambda(t)$ is a diagonal matrix containing the eigenvalues $\lambda_s^m(t)$ and where 
$$
M_{km}(t)=-\tau {\rm Tr }\left(\ell_s^k(t) \dot{r}_s^m(t)\right)\, .
$$
The latter matrix has finite values which further remain finite in the large $\tau$ limit, since we are considering finite-dimensional systems and  $\dot{r}_s^m(t)$ of order $1/\tau$. This implies that there exists a constant $M_{\rm max}$ such that $\|M(t)\|\le M_{\rm max}$ for all $t\ge0$. (We note that the diagonal elements of the matrix $M$ are analogous to the geometric phase in the Hamiltonian case \cite{berry1984, RevModPhys.82.1959, shapere1989geometric}.)
The full propagator for the time evolution of the coefficients as expressed by Eq.~\eqref{coeff-der-vec} is given by the time-order exponential $U(t) = \mathcal{T} \exp\{\int_0^t[\Lambda(v) + \tau^{-1}M(v)] {\rm d}v\}$. This shows that the coefficients ${\bf c}_s(t)$ evolves through a diagonal matrix $\Lambda(t)$ plus an off-diagonal one of order $1/\tau$. Our aim is to show that, in the large $\tau$ limit, all coefficients $c_s^m(t)$ are vanishingly small when compared with the coefficient $c_s^0(t)$, which is enough to prove Eq.~\eqref{ad-theorem}. 

We start by expanding the propagator $U(t)$ into a Dyson series using the matrix $M(t)$ as  a perturbation and the parameter $1/\tau$ as the perturbation strength. The Dyson series reads $U(t)=\sum_{n=0}^\infty V_n$, with 
$$
V_n=\frac{1}{\tau^n}\int_0^t \!\!{\rm d}v_n\int_0^{v_n}\!\!{\rm d}v_{n-1}\dots \int_0^{v_2}\!\!{\rm d}v_1 B(t,v_n)M(v_n)B(v_n,v_{n-1})M(v_{n-1})\dots B(v_2,v_1)M(v_1)B(v_1,0)\, ,
$$
where we have defined
\begin{equation*}
    B(t,t_0) = e^{\int_{t_0}^t \Lambda(v) \mathrm{d}v}  \, .
\end{equation*}
The evolution of the coefficients ${\bf c}_s(t)$ is then given by 
$$
{\bf c}_s(t)=B(t,0){\bf c}_s(0) +\sum_{n=1}^\infty V_n{\bf c}_s(0)\, ,
$$
where we have singled out the zeroth term of the series. From this term, it is clear that the dominant contribution is given by the term $e^{\int_0^t {\rm d}v \lambda_s^0(v)}$ included in $B(t,0)$. We thus want to understand how all the coefficients behave in comparison with this quantity. We take a generic element $c_s^m(t)$ in the vector ${\bf c}_s(t)$ and consider the ratio
\begin{equation}
\label{series-coeff}
\frac{c_s^m(t)}{e^{\int_0^t {\rm d}v \lambda_s^0(v)}}=\frac{e^{\int_0^t {\rm d}v \lambda_s^m(v)}}{e^{\int_0^t {\rm d}v \lambda_s^0(v)}}c_s^m(0)+\sum_{n=1}^\infty \frac{{\bf e }_m\cdot  V_n {\bf c}_s(0)}{e^{\int_0^t {\rm d}v \lambda_s^0(v)}}\, ,
\end{equation}
where ${\bf e}_m$ is the basis vector with $1$ in the $m$th position and $0$ otherwise. We now want to bound each term on the right hand side of the above equation for $m\neq0$. For the first term we find 
$$
\left|\frac{e^{\int_0^t {\rm d}v \lambda_s^m(v)}}{e^{\int_0^t {\rm d}v \lambda_s^0(v)}}c_s^m(0)\right|\le \left|\frac{e^{\int_0^t {\rm d}v \lambda_s^m(v)}}{e^{\int_0^t {\rm d}v \lambda_s^0(v)}}\right|c_{\rm max}\, ,
$$
where $c_{\rm max}$ is such that $|c_s^m(0)|\le c_{\rm max} $, $\forall m$. Considering that we can write the ratio of the exponential as 
$$
\left|\frac{e^{\int_0^t {\rm d}v \lambda_s^m(v)}}{e^{\int_0^t {\rm d}v \lambda_s^0(v)}}\right|=\left|e^{\int_0^t {\rm d}v [\lambda_s^m(v)-\lambda_s^0(v)]}\right|\, ,
$$
and using our finite-gap condition (C2), we find (recalling that  $m\neq0$)
\begin{equation}
\label{first-term}
\left|\frac{e^{\int_0^t {\rm d}v \lambda_s^m(v)}}{e^{\int_0^t {\rm d}v \lambda_s^0(v)}}c_s^m(0)\right|\le e^{-\Delta t}c_{\rm max}\, .
\end{equation}
Now, we consider the generic term in the series of Eq.~\eqref{series-coeff}. Explicitly, we have 
\begin{equation}
\begin{split}
\frac{{\bf e }_m\cdot  V_n {\bf c}_s(0)}{e^{\int_0^t {\rm d}v \lambda_s^0(v)}}&=\frac{1}{\tau^n}\int_0^t \!\!{\rm d}v_n\int_0^{v_n}\!\!{\rm d}v_{n-1}\dots \int_0^{v_2}\!\!{\rm d}v_1\times\\
&\times \frac{  e^{\int_{v_n}^t {\rm d}r_n \lambda_{s}^m(r_n)}{\bf e }_m\cdot  M(v_n)B(v_n,v_{n-1})M(v_{n-1})\dots B(v_2,v_1)M(v_1)B(v_1,0) {\bf c}_s(0)}{e^{\int_0^t {\rm d}v \lambda_s^0(v)}}\, .
\end{split}
\end{equation}
We now split the exponential $e^{\int_0^t {\rm d}v \lambda_s^0(v)}$, which does not depend on the integration variables, into the product 
$$
e^{\int_0^t {\rm d}v \lambda_s^0(v)}=e^{\int_{v_n}^t {\rm d}r_n \lambda_s^0(r_n)}e^{\int_{v_{n-1}}^{v_n} {\rm d}r_{n-1} \lambda_s^0(r_{n-1})}\dots e^{\int_{v_{1}}^{v_2} {\rm d}r_{1} \lambda_s^0(r_{1})}e^{\int_{0}^{v_1} {\rm d}r \lambda_s^0(r)}
$$
and assign each term to the corresponding $B(v_k,v_{k-1})$ and to the first exponential. We can thus write 
\begin{equation}
\begin{split}
\frac{{\bf e }_m\cdot  V_n {\bf c}_s(0)}{e^{\int_0^t {\rm d}v \lambda_s^0(v)}}&=\frac{1}{\tau^n}\int_0^t \!\!{\rm d}v_n\int_0^{v_n}\!\!{\rm d}v_{n-1}\dots \int_0^{v_2}\!\!{\rm d}v_1e^{\int_{v_n}^t {\rm d}r_n [\lambda_{s}^m(r_n)-\lambda_s^0(r_n)]}   \times\\
&\times {\bf e }_m\cdot  M(v_n)\frac{B(v_n,v_{n-1})}{e^{\int_{v_{n-1}}^{v_n} {\rm d}r_{n-1} \lambda_s^0(r_{n-1})}}M(v_{n-1})\dots \frac{B(v_2,v_1)}{e^{\int_{v_{1}}^{v_2} {\rm d}r_{1} \lambda_s^0(r_{1})}}M(v_1)\frac{B(v_1,0) }{e^{\int_0^{v_1} {\rm d}r \lambda_s^0(r)}}{\bf c}_s(0)\, .
\end{split}
\end{equation}
Taking the absolute value of the above quantity, using the Cauchy-Schwarz inequality for the matrix element in the second line, and noticing that $\left\|\frac{B(v_k,v_{k-1})}{e^{\int_{v_{k-1}}^{v_k} {\rm d}r_{k-1} \lambda_s^0(r_{k-1})}}\right\|\le 1$, we find 
$$
\left|\frac{{\bf e }_m\cdot  V_n {\bf c}_s(0)}{e^{\int_0^t {\rm d}v \lambda_s^0(v)}}\right|\le  \frac{\|{\bf  c}_s(0)\| M_{\rm max}^n}{\tau^n}\int_0^t \!\!{\rm d}v_n e^{-\Delta (t-v_n)}  \int_0^{v_n}\!\!{\rm d}v_{n-1}\dots \int_0^{v_2}\!\!{\rm d}v_1 \, ,
$$
where we have exploited our gap condition for the first exponential. Calculating the integrals up to the integration variable  $v_{n-1}$ we find 
$$
\int_0^{v_n}\!\!{\rm d}v_{n-1}\dots \int_0^{v_2}\!\!{\rm d}v_1 =\frac{v_n^{n-1}}{(n-1)!}\le \frac{\tau^{n-1}}{(n-1)!}\, ,
$$
which, when substituted in the bound above, gives 
$$
\left|\frac{{\bf e }_m\cdot  V_n {\bf c}_s(0)}{e^{\int_0^t {\rm d}v \lambda_s^0(v)}}\right|\le  \frac{c_{\rm max} D^2}{\tau \Delta} \left(1-e^{-\Delta t} \right) \frac{M_{\rm max}^n}{(n-1)!}  \, ,
$$
where we further used that $\|{\bf  c}_s(0)\| \le c_{\rm max} D^2$ with $D$ being the dimension of the Hilbert space of the system. Putting all together, this shows that, for $m\neq0$,
$$
\left|\frac{c_s^m(t)}{e^{\int_0^t {\rm d}v \lambda_s^0(v)}}\right|\le c_{\rm max}\left[e^{-t\Delta }+\frac{D^2}{\tau \Delta}(1-e^{-t\Delta})\sum_{n=1}^\infty \frac{M_{\rm max}^n}{(n-1)!}\right]\, ,
$$
which vanishes when introducing the rescaled time $u=t/\tau$ and sending $\tau\to \infty$. Doing a similar manipulation to the one performed above one can instead show that $c_s^0(t)/e^{\int_0^t {\rm d}r \lambda_s^0(r)}$ remains finite in the large-$\tau$ limit. 
The above result implies that $\rho_s(u)\approx c_s^0(u) r_s^0(u)$ for large $\tau$ and thus that 
$$
\lim_{\tau\to\infty}\varrho_s(u):=\lim_{\tau\to\infty}\frac{\rho_s(u)}{{\rm Tr}(\rho_s(u))}=r_s^0(u)\, ,
$$
where we defined $r_s^0(u)$ to be unit trace. \\

We conclude this section by briefly discussing the case of a degenerate dominant eigenvalue $\lambda_s^0(s)$. Even with such a degeneracy, it is still possible to do the same steps done above to show that all the coefficients $c_s^m$, which are not associated with the dominant eigenvalue, would be vanishing in the large-$\tau$ limit. That is, the quantum state $\varrho_s(u)$ would only contain eigenmatrices $r_s^m(u)$ related to the dominant eigenvalue. As such, even though the form of $\varrho_s(u)$ would not be completely determined by the above calculation, one can still conclude that $\mathcal{L}_s(u)[\varrho_s(u)]=\lambda_s^0(u)\varrho_s(u)$, which is all that is needed to arrive at Eq.~\eqref{theta-adiabatic}.

\section{III. Observable rates and probability functional for time-histories}
Due to adiabatic character of the dynamics, the system essentially spends an infinite amount of time in each of the rescaled  times $u=t/\tau$. For each of these times, we can thus define a time-averaged value of the observable $q(u)$, which represents an instantaneous rate for the observable $Q(\tau)$. As a special case of the derivation we present below, the latter can indeed be written as $Q(\tau)=\tau\int_0^1 q(u) {\rm d}u$. In this Section, we show how the functional providing the probability of any time-history, or path, $\{q(u)\}$ can be derived. 

The first step consists in showing that the moment generating function for the time-histories $\{q(u)\}$ can be defined by considering a time-dependent {\it path} for the conjugate field $s(u)$, which only varies on the slow timescale $u$. To this end, we start defining the quantity
\begin{equation}
\hat{Q}(\tau)=\sum_j \int_0^\tau s(v/\tau) f_j(v/\tau) {\rm d} n_j(v)\, ,
\end{equation}
where the factor $1/\tau$ in the arguments of the functions $s$ and $f_j$ explicitly accounts for the fact that these functions vary on the slow timescale. To proceed, we now divide the total evolution time $\tau$ into $M$ time-intervals so that we can write
$$
\hat{Q}(\tau) =\sum_{k=1}^M \sum_j \int_{\frac{\tau}{M}(k-1)}^{\frac{\tau}{M}k} s(v/\tau) f_j(v/\tau) {\rm d}n_j(v)\, .
$$
For $M$ large enough, the functions $s$ and $f_j$ are essentially constant inside each time interval so that we have 
$$
\hat{Q}(\tau) \approx \sum_{k=1}^M \sum_j s(k/M)f_j(k/M)  \int_{\frac{\tau}{M}(k-1)}^{\frac{\tau}{M}k}  {\rm d}n_j(v)=\frac{\tau}{M}\sum_{k=1}^M  s(k/M) \left[\sum_j f_j(k/M) \frac{1}{\tau/M} \int_{\frac{\tau}{M}(k-1)}^{\frac{\tau}{M}k}  {\rm d}n_j(v)\right]\, ,
$$
where for the second equality we only  multiplied and divided by the factor $\tau/M$. The term inside the square bracket is essentially the average rate of the considered observable at the rescaled time $k/M$, which we call $q(k/M)$. With such a definition, we can write 
$$
\hat{Q}(\tau)\approx \frac{\tau}{M}\sum_{k=1}^M s(k/M) q(k/M)\, .
$$
Taking the limit $M\to\infty$ we find 
$$
\hat{Q}(\tau)=\tau \int_0^1 s(u) q(u) {\rm d}u\, .
$$
The above expression thus shows that a time-dependent field $s(u)$ is conjugated to the time-history of the rate $\{q(u)\}$, whose moment generating function can thus be written as $Z_{\{s(u)\}}=\mathbb{E}[e^{-\hat{Q}(\tau)}]$. 

The second step to arrive at a probability function for time-histories is to recognize that the tilted operator for $Z_{\{s(u)\}}$ is exactly the one presented in Eq.~\eqref{eq:tilted_Lindbladian} but with a time-dependent $s$, $\mathcal{L}_{s(u)}(u)$ varying on the slow timescale $u$. This thus also implies that our adiabatic theorem is also valid for this tilted generator and that we can write the scaled cumulant generating functional for time-histories as 
$$
\Theta[\{s(u)\}]=\int_0^1 \lambda_{s(u)}^0(u) {\rm d}u\, ,
$$
where $\lambda_{s(u)}^0(u)$ is the dominant eigenvalue of $\mathcal{L}_{s(u)}(u)$.

With the scaled cumulant generating function at hand, we can calculate the probability functional for time-histories as its Legendre-Fenchel transform. That is, the large deviation rate function for $\{q(u)\}$ is defined as 
$$
\varphi[\{q(u)\}]=\sup_{\{s(u)\}}\left[-\int_0^1 s(u)q(u) {\rm d}u-\Theta[\{s(u)\}]\right]\, .
$$
Taking the functional derivative with respect to $\delta s(r)$, we find 
$$
\frac{\delta}{\delta s(r)}\left[-\int_0^1 s(u)q(u) {\rm d}u-\Theta[\{s(u)\}]\right]=-q(r)-\frac{\delta \lambda_{s(r)}(r)}{\delta s(r)}.
$$
Then, assuming that it is possible to find the suitable path $s^*(r)$ such that the above equality can be set to zero, we find 
$$
q(r)=-\frac{\delta \lambda_{s^*(r)}(r)}{\delta s^*(r)}\, .
$$
Substituting into the equation for $\varphi$, we obtain 
$$
\varphi[\{q(u)\}]=\int_0^1 \left[s^*(u) \frac{\delta \lambda_{s^*(u)}(u)}{\delta s^*(u)}- \lambda_{s^*(u)}(u) \right]{\rm d}u \, .
$$
For each $u$, the term inside the integral corresponds to the Legendre-Fenchel transform of the instantaneous dominant eigenvalue, calculated in $q(u)$, so that we finally have 
$$
\varphi[\{q(u)\}]=\int_0^1 \mathcal{I}(q(u),u) {\rm d} u\, .
$$

\subsection{Contraction to the full counting statistics of the time-integrated observable}

In this Section, we explicitly show how to go from the functional over time-histories $\varphi$ to the large deviation rate function $I$ for the time-integrated observable $Q(\tau)=\tau\int_0^\tau q(u){\rm d}u$. This can be done via the contraction principle given that the observable $Q$ is a function of the different time-histories.
Due to the contraction principle, we can define, as done in the main text, 
$$
I(Q^*/\tau)=\min_{\forall \{q(u)\}: \int_0^1q(u){\rm d}u=Q^*/\tau}\left[\int_0^1 \mathcal{I}(q(u),u){\rm d}u\right]\, .
$$
To perform the constrained minimization we introduce the Lagrange multiplier $\mu$ and construct the functional
$$
Y[\{q(u)\},\mu]=\int_0^1 \mathcal{I}(q(u),u) {\rm d}u -\mu \left[\int_0^1 q(u){\rm d}u-Q^*/\tau\right] \, .
$$
We then take the functional derivative of $Y$ with respect to $\delta q(r)$ to find
\begin{equation}
\label{der_q(r)}
\frac{\delta Y}{\delta q(r)}=\frac{\delta \mathcal{I}(q(r),r)}{\delta q(r)}-\mu\, .
\end{equation}
We assume that it is possible to find the path $q^*(r)$, which depends on $\mu$, such that the above quantity can be made equal to zero. Then, to fix the value of $\mu=\mu^*$, we integrate $q^*(r)$ and insist that 
$$
\int_0^1 q^*(r) {\rm d}r =Q^*/\tau\, .
$$
We can then substitute this into the functional to obtain 
$$
Y({\{q^*(u)\}},\mu^*)=\int_0^1 \mathcal{I}(q^*(r),r) {\rm d}r\, .
$$
The task now is to find a convenient expression for the large deviation rate function  $\mathcal{I}(q^*(r),r)$. To this end, we note that this is the Legendre-Fenchel transform of the instantaneous dominant eigenvalue $\lambda_\mu(r)$. This means that we also have the inverse relation 
$$
\lambda_{\mu^*}(r)=-\inf_{\forall q(r)} \left[\mathcal{I}(q(r),r)+\mu^* q(r)\right]. 
$$
Performing the minimization we find the same relation as in Eq.~\eqref{der_q(r)}, i.e.,  $\delta\mathcal{I}(q(r),r)/\delta q^*(r)=\mu^*$. Substituting this in the equation for $\lambda_\mu^*(r)$ we obtain the relation 
$$
\lambda_{\mu^*}(r)=-\mathcal{I}(q^*(r),r)-\mu^* q^*(r)\, .
$$
Solving this for  $\mathcal{I}(q^*(r),r)$ and substituting the in the $Y$ functional we find
$$
I(Q^*/\tau)=Y(\{q^*(u),\mu^*\}=-\mu^*\int_0^1 q^*(u) {\rm d} u -\int_0^1 \lambda_{\mu^*}(u){\rm d}u =-\mu^* Q^*/\tau -\theta_{\mu^*}^{\rm ad}\, ,
$$
which is exactly the Legendre-Fenchel transform of $\theta_\mu^{\rm ad}$ as it should be.

\section{IV. Quantum Doob Transform for time-dependent generators}
In this Section, we discuss how it is possible to define an auxiliary process that gives a desired time-history for the observable as its typical one. This is achieved by exploiting ideas put forward in Refs.~\cite{PhysRevLett.104.160601,PhysRevA.98.010103} in the context of quantum  generalizations of the \textit{Doob transform} introduced in Ref.~\cite{jack2010}. 

We start by considering the evolution under the deformed dynamical generator and Trotterize this in the slow time-scale. We can thus write 
$$
\rho_s(\tau)=\mathcal{T} e^{\tau \int_0^{1}  \mathcal{L}_{s(u)}(u){\rm d} u} [\rho(0)]\approx \prod\limits_{k=1}^M e^{\tau \mathcal{L}_{s(k/M)}(k/M){\rm d} u}[\rho(0)]  \tc
$$
where we considered the same partitioning of the $\tau$ exploited in Section III. For a given path $\{s(u)\}$ this deformed generator enhances the probability of observing as typical time-history the one given by $q(u)=-\delta \Theta/\delta s(u)$. However, the above evolution is not a physical dynamics. To obtain a suitable, we can proceed by applying at each rescaled time the time-independent Doob transform introduced in Refs.~\cite{PhysRevLett.104.160601,PhysRevA.98.010103}. Essentially, we want to find a generalized rotation of the deformed generator at each rescaled time $u$, $\mathcal{L}_{s(u)}(u)$, such that the latter becomes a well-defined quantum map. Following \cite{PhysRevLett.104.160601}, this is achieved by exploiting the left dominant eigenmatrix of the deformed generator and defined the auxiliary dynamics via the dynamical generator 
\begin{equation*}
\mathcal{L}^{\rm A}(u)[\rho]=-i[H^{\rm A}(u),\rho]+\sum_j\left[J_j^{\rm A}(u)\rho (J_j^{\rm A}(u))^\dagger -\frac{1}{2}\left\{(J_j^{\rm A}(u))^\dagger J_j^{\rm A}(u),\rho\right\}\right] \tc
\end{equation*}
with Hamiltonian and jump operators given in Eq.~\eqref{Doob_H} of the main text. The dynamics resulting from the generator $\mathcal{L}^{\rm A}(u)$ is still an adiabatic dynamics by construction and such that the typical path is the one given by $q(u)=-\delta \Theta/\delta s(u)$ as desired.

\section{V. Master equation of the two-qubit system}
In this Section, we provide details on the quantum master equation for the two-qubit system coupled to two different thermal baths, a hot one and a cold one.

We consider the Hamiltonian 
\begin{equation}
    H = \omega(\sigma_e^{\text{hot}} + \sigma_e^{\text{cold}}) + \Omega(\sigma_+^{\text{hot}}\sigma_-^{\text{cold}} + \sigma_-^{\text{hot}}\sigma_+^{\text{cold}}) \tc
\end{equation}
where $\Omega$ is the interaction strength, $x^{\text{hot}}=x \otimes \1$ and $x^{\text{cold}}=\1 \otimes x$  concern the qubit in contact with the hot bath and the cold one, respectively. The operator $\sigma_e$ is the project onto the excited state $n=\ket{e}\bra{e}$ while $\sigma_-=\ket{g}\bra{e}$ and $\sigma_+=\sigma_-^\dagger$. Under the assumption that the qubits are interacting with a hot and a cold thermal bath, we find that, under a weak-coupling assumption, the dynamics of the state of the system is  described by the Lindblad generator 
\begin{equation}
    \dot{\rho}(t)= -i[H, \rho(t)] + \sum_{j}\left(J_j \rho(t) J_j^\dag -\frac{1}{2}\{J_i^\dag J_i, \rho(t) \} \right) \tc 
\end{equation}
where we have the jump operators
\begin{align}
    J_{01}^b &= \sqrt{\gamma}\sqrt{ \left(\frac{\omega-\Omega}{\omega}\right)^3\left(1 + \frac{1}{e^{\beta_b (\omega-\Omega)} - 1}\right)}\ketbra{\epsilon_0}{\epsilon_1} \,,\quad J_{10}^b = \sqrt{\gamma}\sqrt{ \left(\frac{\omega-\Omega}{\omega}\right)^3\frac{1}{e^{\beta_b (\omega-\Omega)} - 1}}\ketbra{\epsilon_1}{\epsilon_0} \,, \\
    J_{02}^b &= \sqrt{\gamma}\sqrt{ \left(\frac{\omega+\Omega}{\omega}\right)^3\left(1 + \frac{1}{e^{\beta_b (\omega+\Omega)} - 1}\right)}\ketbra{\epsilon_0}{\epsilon_2}\,, \quad J_{20}^b = \sqrt{\gamma}\sqrt{ \left(\frac{\omega +\Omega}{\omega}\right)^3\frac{1}{e^{\beta_b (\omega+\Omega)} - 1}}\ketbra{\epsilon_2}{\epsilon_0} \,, \\
    J_{13}^b &= \sqrt{\gamma}\sqrt{ \left(\frac{\omega +\Omega}{\omega}\right)^3\left(1 + \frac{1}{e^{\beta_b (\omega+\Omega)} - 1}\right)}\ketbra{\epsilon_1}{\epsilon_3}\,,  \quad J_{31}^b = \sqrt{\gamma}\sqrt{ \left(\frac{\omega+\Omega}{\omega}\right)^3\frac{1}{e^{\beta_b (\omega+\Omega)} - 1}}\ketbra{\epsilon_3}{\epsilon_1} \,,\\
    J_{23}^b &= \sqrt{\gamma}\sqrt{ \left(\frac{\omega-\Omega}{\omega}\right)^3\left(1 + \frac{1}{e^{\beta_b (\omega-\Omega)} - 1}\right)}\ketbra{\epsilon_2}{\epsilon_3}\,, \quad J_{32}^b = \sqrt{\gamma}\sqrt{ \left(\frac{\omega-\Omega}{\omega}\right)^3\frac{1}{e^{\beta_b (\omega-\Omega)} - 1}}\ketbra{\epsilon_3}{\epsilon_2} \,.
\end{align}
Here, $\gamma$ is a rate and $\beta_b$, with $b=\text{hot},\text{cold}$, is the inverse temperature of the bath. The vectors $\ket{\epsilon_i}$ are the eigenstates of the Hamiltonian $H$, given by 
\begin{equation}
    \begin{split}
        |\epsilon_0  \rangle = |gg\rangle \,, \quad|\epsilon_{1,2} \rangle = \frac{1}{\sqrt{2}}\left(|ge\rangle \mp |eg\rangle\right) \,, \quad  |\epsilon_3 \rangle = |ee\rangle \tc
    \end{split}
\end{equation}
where $|ij\rangle = |i\rangle\otimes |j\rangle$, associated with the energies  $\epsilon_0=0$, $\epsilon_{1,2}=\omega\mp \Omega$, and $\epsilon_3 = 2\omega$. In addition to these terms, we further consider phenomenologically a laser driving given by $H_S=H+H_\text{laser}$ with $H_{\text {laser}}=g(\sigma_x^{\text{hot}} + \sigma_x^{\text{cold}})$.

\end{document}